\documentclass[aps,prd,reprint,superscriptaddress,nofootinbib,showkeys,amsmath,amssymb,floatfix]{revtex4-2}

\usepackage{amssymb}
\usepackage{amsmath}
\usepackage{bm}
\usepackage[T1]{fontenc}
\usepackage{graphicx}
\usepackage[colorlinks]{hyperref}

\begin{document}

\title{Spatial Qubit Entanglement Witness for Quantum Natured Gravity}

\author{Bin Yi}
\affiliation{Institute of Fundamental and Frontier Sciences, University of Electronic Science and Technology of China, Chengdu 611731, China}
\affiliation{Key Laboratory of Quantum Physics and Photonic Quantum Information, Ministry of Education, University of Electronic Science and Technology of China, Chengdu 611731, China}
\affiliation{Department of Physics and Astronomy, University College London, Gower Street, WC1E 6BT London, United Kingdom}

\author{Urbasi Sinha}
\affiliation{Raman Research Institute, C.~V.~Raman Avenue, Sadashivanagar, Bengaluru, Karnataka 560080, India}
\affiliation{Department of Physics and Astronomy, University of Calgary, Alberta T2N 1N4, Canada}

\author{Dipankar Home}
\affiliation{Center for Astroparticle Physics and Space Science (CAPSS), Bose Institute, Kolkata 700~091, India}

\author{Anupam Mazumdar}
\affiliation{University of Groningen, PO Box 72, 9700 Groningen, Netherlands}
\affiliation{Van Swinderen Institute, University of Groningen, 9747 AG Groningen, Netherlands}

\author{Sougato Bose}
\affiliation{Department of Physics and Astronomy, University College London, Gower Street, WC1E 6BT London, United Kingdom}

\date{\today}

\begin{abstract}
Witnessing the quantum nature of gravity through the entanglement of two masses has recently been proposed. Both non-Gaussian and Gaussian protocols have been shown to be useful for this purpose. In non-Gaussian protocols the complete $1/r$ gravitational interaction can in principle be utilized for entanglement generation, as long as quantum superpositions of a spatial splitting $\Delta x$ comparable to the distance $d$ between two masses can be generated. In practice, however, all non-Gaussian protocols proposed to date rely on embedded qubits -- typically spins -- to read out the entanglement: spin correlations are used to witness the entanglement developed between the masses during interferometry. Here we show that if superpositions of distinct spatially localized states of each mass can be created, then a spinless version of the non-Gaussian protocol is possible: simple position correlation measurements alone can yield a spatial qubit witness of the entanglement between the masses. We find that the principal new requirement for the viability of this protocol is a position squeezing of the wavefunction by approximately seven orders of magnitude, applied at a specific stage of the experiment.
\end{abstract}


\maketitle

\section{Introduction}
The quantumness of gravity is an open question due to a lack of empirical evidence. A lot of research is performed in the setting of semi-classical gravity, in which matter and non-gravitational fields are treated quantum mechanically, while gravity is treated classically. A substantial community argues that gravity can be classical as quantum physics break down at macroscopic level, where gravitational effects become evident\cite{penrose1996gravity,diosi1989models,bassi2013models,bassi2017gravitational}. Various proposals have been made to avoid quantizing gravity, at the cost of introducing extra stochastic terms\cite{kafri2014classical,oppenheim2022constraints} -- but these are not ruled out by any current experiment such as measuring forces precisely. Therefore, testing the quantum nature of gravity experimentally is an open problem. Even if its quantumness is accepted from the point of view of several existing successful theories \cite{oriti2009approaches,kiefer2006quantum}, its \emph{verification} is still open.

In 2017, the authors of Ref.\cite{bose2017spin} proposed a protocol to test whether the nature of gravity is quantum. Two spatial superposition states of masses \emph{cannot entangle} via a classical channel \cite{bennett1999quantum} (see also \cite{marletto2017gravitationally}). Locality in quantum field theory circumvents the non-local interaction between the two superposed test masses\cite{marshman2020locality}. Entanglement between the masses can only be generated through local operations and \emph{quantum} communication; in fact, quantum communication is \emph{necessary} for operator-valued interactions \cite{bose2022mechanism}, which in turn are necessary for the coherent interaction that generates entanglement. Once quantum communication is proven through the witnessing of gravitationally generated entanglement, this unambiguously certifies the presence of off-shell/virtual gravitons as the only way to obtain a continuous deterministic operator-valued interaction\cite{marshman2020locality,bose2022mechanism}. Alternatively, the witnessed entanglement can also be regarded as evidencing the quantum superposition of geometries inherent in superposition states of each mass \cite{christodoulou2019possibility,christodoulou2022locally}. Moreover, from logical arguments, the quantum nature of the Newtonian component of the gravitational field automatically has a bearing on the quantum nature of other components \cite{belenchia2018quantum,danielson2022gravitationally,carney2022newton}.

The original proposal exploits spin-embedded masses\cite{bose2017spin}. This methodology was adopted because there was no \emph{known} mechanism for implementing a beam splitter for nano-crystal masses, so that a Mach--Zehnder interferometer cannot be implemented. Stern--Gerlach interferometers (SGIs) are used to prepare the spatial superpositions which depend on embedded spins so that the motional and spin degrees of freedom are entangled. The test masses then freely evolve subject to gravitational interaction. At the end of the proposed protocol, the Stern--Gerlach apparatus is exploited again to bring the superposition components back to the center. Spin correlations between the test masses then evidence the entanglement generated through gravitational interaction during propagation. Alternatively, it has also been proposed to witness the gravitational entanglement growth between two initially delocalized Gaussian states by position and momentum correlations \cite{qvarfort2020mesoscopic, krisnanda2020observable, weiss2021large}. However, the Gaussian approximation necessitates that the masses be kept much farther than their wavepacket delocalizations. In contrast, a two-qubit witness for entanglement is applicable in regimes where the masses are brought as close to each other as their delocalization (spatial superposition scale), so that the entanglement itself has a reasonably fast growth rate ($\sim 1$ Hz) even for smaller (micron-sized) masses.

\emph{The motivation for a spinless protocol.} The spin-based SGI scheme faces two intrinsic obstacles. First, completing the SGI requires an exact overlap to be attained in both position and momentum of the wavepackets in the two interferometric arms \cite{englert1988spin,Folman2013}, a feat which has only very recently been achieved for small atomic interferometers \cite{margalit2020realization} and remains formidable at the $\sim 10^{-15}\,$kg, $10\,\mu$m scale required for the BMV test. Second, the embedded spin introduces an additional decoherence channel -- already characterized in detail for nitrogen-vacancy electron spins coupled to nuclear spin baths \cite{xing2016electron} -- and calls for elaborate dynamical-decoupling sequences \cite{bar2013solid,wood2022spin}, complicating the interferometry further. Although exceptionally long spin coherences have been demonstrated in qubit platforms \cite{zhang2025ultralong}, it is worthwhile to ask whether the entanglement can be witnessed \emph{without} spins altogether. We thus pursue the alternative of using position measurements themselves to infer the states of \emph{spatial qubits} -- which have been called Young qubits in photonic systems \cite{taguchi2008measurement,ghosh2018spatially}, the methodology having been experimentally demonstrated in the photonic setting in Ref.\cite{ghosh2018spatially} -- and use this as a witness of gravitational entanglement. We find that, aside from the usual coherence requirements that must be met to maintain the spatial superpositions \cite{bose2017spin,van2020quantum} -- unavoidable in any scheme -- a spatial qubit witnessing of gravitational entanglement is possible \emph{provided a challenging squeezing requirement can be met}.

\section{Background: spin-based gravitational entanglement witnessing}
Ref.\cite{bose2017spin} presents an unambiguous witness of quantum natured gravity. The scheme consists of two spin-embedded test masses initially in spatially localized states, say in respective traps. The test masses pass through a set of Stern--Gerlach (SG) interferometers so that the spatial degree of freedom entangles with the spin degree of freedom to prepare
\begin{equation}
 |\psi_0\rangle_j=|L,\uparrow\rangle_j+|R,\downarrow\rangle_j\quad(j=1,2)
\end{equation}
with states $|L\rangle$ and $|R\rangle$ being separated by $d$, and the distance between the midpoints of the two superpositions ($\sim$ the initial separation between the masses) being $D$. Next, let the coherent superposition state propagate for a time $\tau$ by switching off the magnetic field of the SG apparatus. If gravity were quantum, gravitational interaction can induce, through relative phases among the superposition components, an entanglement between the masses (classical gravity as mediator would not give an operator-valued interaction \cite{bose2022mechanism}, and hence will be unable to entangle the two masses). The final step refocuses the SG apparatus $|L,\uparrow\rangle_j\rightarrow |C,\uparrow\rangle_j$, $|R,\downarrow\rangle_j \rightarrow |C,\downarrow\rangle_j$ to bring the spatial superposition in each interferometer back to the center, denoted $C$, so that the final spin state reads
\begin{eqnarray}
\frac{e^{i\phi}}{\sqrt2} &&[|\uparrow\rangle_1\tfrac{1}{\sqrt2}(|\uparrow\rangle_2+e^{i\Delta \phi_{LR}}|\downarrow\rangle_2)\nonumber\\
&&+|\downarrow\rangle_1\tfrac{1}{\sqrt2}(e^{i\Delta \phi_{RL}}|\uparrow\rangle_2+|\downarrow\rangle_2)]\label{bose}
\end{eqnarray}
where $\phi=\frac{Gm_1m_2}{\hbar D}\tau,\ \Delta\phi_{LR}=\frac{Gm_1m_2}{\hbar (D+d)}\tau-\phi,\ \Delta\phi_{RL}=\frac{Gm_1m_2}{\hbar (D-d)}\tau-\phi$. By measuring spin correlations one can verify the entanglement induced during the propagation time $\tau$. That can only arise from the exchange of quantum coherent mediators. If gravity is the only interaction present, one can then conclude that gravity is quantum.

To produce an observable relative phase, the original proposal considers massive objects with $m\sim10^{-14}\,$kg, restricted by the minimum distance between the two masses. Casimir--Polder force becomes dominant over gravitational interaction at micron-scale separations, requiring $D-d\sim 100\,\mu$m. A revised scheme based on Casimir screening \cite{van2020quantum} relaxes the parameters to $D\sim 47\,\mu$m, $d\sim 23\,\mu$m by placing a conducting plate, acting as a \emph{Faraday cage}, between the test masses. Testing the spin entanglement witness also requires a delicate balancing of magnetic field gradients to bring the superposition components back to the center. Completing the SGI -- with exact phase-space overlap of the two arms -- and preserving spin coherence throughout are the two principal challenges of the spin-based scheme.

\section{Background: massive spatial qubit}
We now investigate the viability of testing the quantum nature of gravity with the recently developed methodology of \emph{massive spatial qubits}\cite{yi2021massive}. This approach treats freely evolving spatially superposed masses as qubits, which does not call for spins. Therefore it may relax the criteria for witnessing the quantum aspect of gravity. The two spatial superposition components of a freely propagating particle, $|L\rangle$ and $|R\rangle$, are identified as the basis of the qubit encoding. To read out the encoded information one performs Pauli measurements by placing spatial detectors at particular locations. The Pauli-$z$ measurement $\sigma_z = |L\rangle\langle L|-|R\rangle\langle R|$ reads the probability amplitude of the encoded spatial qubit state, which has to be performed \emph{before} $|L\rangle$ and $|R\rangle$ have spread so much as to be confused with each other (i.e.\ spread to about $\sim d$). On the other hand, Pauli-$x$ measurements project onto the states $\frac{1}{\sqrt2}(|L\rangle\pm|R\rangle)$ and Pauli-$y$ measurements project onto $\frac{1}{\sqrt2}(|L\rangle\pm i|R\rangle)$, which are only discernible in the interference plane. They are performed \emph{after} each of $|L\rangle$ and $|R\rangle$ have spread to a length $d$ so that they can overlap. At the interference plane, the probability of detecting the test mass at position $x$ is given by
\begin{equation}
P(x)=|\langle\psi|k_x\rangle|^2 \propto \left| \exp\left[\tfrac{ik_xd}{2}-k_x^2\sigma_d^2\right]\langle\theta|\psi\rangle \right|^2,
\end{equation}
where $|\theta\rangle=|0\rangle+|1\rangle e^{i\theta}$, $\theta=k_xd$, and $k_x=\frac{xm}{\hbar t^{x,y}_{\rm meas}}$ is the transverse wavevector. Projection of the spatial qubit state along $\sigma_{x\pm}$ ($\sigma_{y\pm}$) can then be implemented by placing spatial detectors at $\theta=0,\pi$ ($-\pi/2,\pi/2$). Thus the Pauli-$x,y$ and Pauli-$z$ measurements would normally have to be performed at different times $t^z_{\rm meas}$ (before spreading) and $t^{x,y}_{\rm meas}$ (after spreading).

\section{\label{sec:protocol}Spatial qubit methodology for witnessing gravitational entanglement}
\begin{figure}[h!]
\includegraphics[width=0.95\columnwidth]{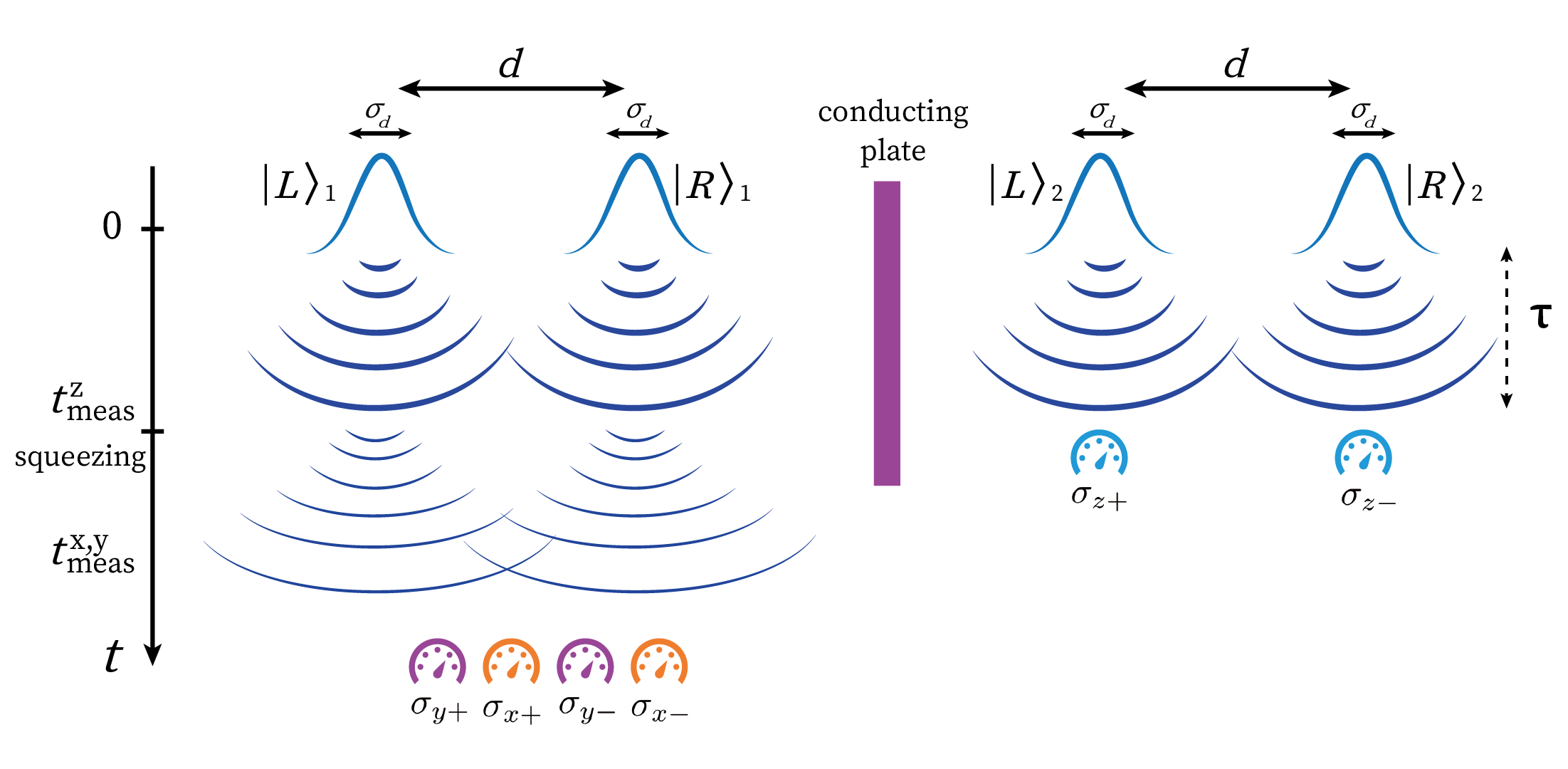}
\caption{\footnotesize \textbf{Testing quantumness of gravity with massive spatial qubits.} Gravitational interaction induces entanglement that can be deduced from spatial qubit Pauli measurements. To evaluate the entanglement witness we conduct spatial qubit measurements on both masses and evaluate the correlations. For example, in the figure, Pauli-$X$ and Pauli-$Y$ measurements on test mass 1 and Pauli-$Z$ measurement on test mass 2 are depicted. An additional conducting plate is inserted between the masses for electromagnetic shielding.}\label{fig2}
\end{figure}

To apply the massive spatial qubit methodology for witnessing gravitational entanglement, we consider two test masses $m_1,m_2$, each prepared in a spatial superposition of two well-separated Gaussian states $|L\rangle$ and $|R\rangle$, placed adjacent to each other and freely evolving under their own Hamiltonian as well as under their mutual gravitational interaction. The schematic is shown in Fig.\ref{fig2}. The propagation of each individual wavepacket is modeled as the spreading of a Gaussian wavepacket due to free evolution. Each spatial superposition $\frac{1}{\sqrt2}(|L\rangle+|R\rangle)$ of a test mass can be treated as a state of a qubit with the two qubit states identified with $|L\rangle$ and $|R\rangle$.

Any entanglement induced during the propagation process can be witnessed by the correlations between two spatial qubits. All of Pauli-$x,y$ and $z$ measurements are involved in obtaining these correlations. Note that due to the finite size of spatial detectors, the measurements performed are approximate Pauli measurements. The detector size smearing can reduce the efficacy of the entanglement measurements; nonetheless optical experiments on spatial qubits show that one can still derive very good visibilities with proper care taken on the parameters \cite{ghosh2022direct}. The positivity of the resulting effective witness on all separable states is established in Appendix~\ref{app:witness}.

We now address in turn the key requirements and constraints of the spatial qubit protocol: the witness construction and induced phase, the squeezing operation needed to align the measurement times, the Casimir-screening geometry, the decoherence budget, and the broadband force-noise constraint.

\paragraph*{Witnessing the quantum nature of gravity.}
If quantum gravitational interaction is the only source of interaction, the relative phase induced among the superposition components reads
\begin{eqnarray}
|\psi(t=0)\rangle&=&\tfrac{1}{\sqrt2}(|L\rangle_1+|R\rangle_1)\,\tfrac{1}{\sqrt2}(|L\rangle_2+|R\rangle_2),\label{initialstate}\\
|\psi(t=\tau)\rangle&=&\tfrac{e^{i\phi}}{\sqrt2}\big[|L\rangle_1\tfrac{1}{\sqrt2}(|L\rangle_2+e^{i\Delta \phi_{LR}}|R\rangle_2)\nonumber\\
&&+|R\rangle_1\tfrac{1}{\sqrt2}(e^{i\Delta \phi_{RL}}|L\rangle_2+|R\rangle_2)\big].\label{ent-state}
\end{eqnarray}
The creation of a superposition of two localized states (a highly non-Gaussian state) is a \emph{prerequisite} for being able to use spatial qubits. Several routes -- both spinful and spinless -- exist for preparing the required $\frac{1}{\sqrt 2}(|L\rangle+|R\rangle)$ at the relevant mass and splitting scales; we summarize these in Appendix~\ref{app:initstate}. Here we concentrate on \emph{how} to read out spatial qubits in the absence of (in this context impractical) beam-splitter elements, and thereby witness the entanglement of two masses without resorting to spins.

\paragraph{Squeezing requirement.}
The masses have to entangle sufficiently first. This inevitably requires a time $\tau$ over which the gravitational interaction acts between them; this is the earliest time \emph{any} measurement can occur, as we want to measure the two masses after they are entangled. Thus we set $t^z_{\rm meas}=\tau$, ensuring that hardly any spreading of the wavepackets $|L\rangle$ and $|R\rangle$ happens in time $\tau$. However, we want to measure Pauli-$x,y$ on the \emph{same} state, i.e.\ the state evolved up to time $\tau$, since the gravitational interaction cannot be switched off. This demands that we must also measure Pauli-$x,y$ at (or very nearly) the same time. So we need to induce some process at $\tau$ so that the wavepackets immediately spread to $\sim d$ and overlap to interfere. This requires the application of an additional squeezing operator to the masses immediately after time $\tau$. This localizes the wavefunctions so much (each of $|L\rangle$ and $|R\rangle$ is now highly squeezed) that in subsequent free evolution for $t^{x,y}_{\rm meas}-t^z_{\rm meas}$ they spread rapidly to overlap and interfere. The extra squeezing at $\tau$ thus enables us to meet the criterion $t^{x,y}_{\rm meas}\sim t^z_{\rm meas}=\tau$, so that both the Pauli-$z$ and the Pauli-$x,y$ measurements are essentially performed on the same entangled state of the two masses.

The spread of a wavepacket scales approximately inversely with its initial width $\sigma_d(0)$: $\sigma(t)\sim\frac{\hbar t}{m \sigma_d(0)}$. Application of a squeezing operator at $\tau$ reduces $\sigma_d(0)$ and speeds up subsequent spreading. The technique can be implemented by passing the test masses through local harmonic potentials with sudden frequency jumps: $n$ sudden switches between frequencies $\omega_1$ and $\omega_2$, with a quarter period of harmonic evolution at each frequency, squeezes the wavepackets spatially by a factor $(\omega_1/\omega_2)^n$ \cite{janszky1992strong,rashid2016experimental,liu2019photon}. This procedure inevitably consumes a time $t_{\rm squeeze}=n(\pi/2\omega_1+\pi/2\omega_2)$, which must satisfy
\begin{equation}
t_{\rm squeeze}\ll t^{x,y}_{\rm meas}\approx t^z_{\rm meas}=\tau.
\end{equation}
Squeezing thus places significant demands on this protocol. The detailed hardware requirements -- diamagnetic trapping, ultra-low dissipation rates, and the corresponding constraints on temperature and pressure -- are discussed in Appendix~\ref{app:squeezing}.

\paragraph{Casimir screening imposed constraints.}
To shield the system from unwanted electromagnetic interaction, one places a conducting plate between the test masses to act as a \emph{Faraday cage}. For simplicity we assume that the placement of a perfect conductor completely blocks the Casimir interaction between the test masses. The Casimir screening scheme introduces an additional Casimir force between the plate and the test masses, which becomes the dominant force at very small scale and places constraints on the system. The screen is closest to one of the components of the superposition ($|R\rangle$ for the mass to the left of the screen and $|L\rangle$ for the mass to the right). We do not want this component to be pulled so close to the screen that it prevents the overlap between the two components of each mass during the Pauli-$x,y$ measurement. Thus we require that the spread of the wavepackets (which we \emph{exploit} in our scheme) dominates over the displacement due to Casimir force, placing a new constraint on the minimum separation of a mass and the conducting plate. The Casimir force between a plate and a sphere is given by\cite{casimir1948influence}
\begin{equation}
F_{ca}=-\frac{3\hbar c}{2\pi}\left(\frac{\epsilon-1}{\epsilon+2}\right)\frac{R^3}{s^5},
\end{equation}
where $\epsilon$ is the dielectric constant of the test masses, $R$ is the radius of the test masses, and $s=\frac{D-d}{2}$ is the mass--plate separation. Requiring that the wavepacket spread dominates the Casimir-induced displacement by at least one order of magnitude, $0.1\sigma_d(t_1)\geq D_{ca}$, with $D_{ca}=-\frac{9\hbar c}{16\pi^2}\left(\frac{\epsilon-1}{\epsilon+2}\right)\frac{1}{\mu s^5}$ (where $\mu$ is the mass density) \cite{van2020quantum}, and using $\sigma_d(t_1)\sim d \sim s$, the constraint is satisfied for $\mu\sim 3\times10^3\,$kg/m$^3$, $\epsilon=5.7$, and a minimum plate--mass separation $s\sim 12\,\mu$m.

\paragraph{Induced phase.}
With the same mechanism as in \cite{bose2017spin} we can prepare the initial state of Eq.\eqref{initialstate}. Taking $D\sim 40\,\mu$m, $d\sim 10\,\mu$m, for mass $m\sim10^{-15}\,$kg, gravitational interaction, if quantum, would induce a relative phase $\Delta \phi \sim 1\times10^{-2}\,$rad after $\tau\sim 3\,$s of entangling time. To certify the induced relative phase we estimate the entanglement witness \cite{chevalier2020witnessing}
\begin{equation}
\tilde{W}=I\otimes I-\tilde{\sigma}_x\otimes \tilde{\sigma}_x-\sigma_z \otimes \tilde{\sigma}_y -\tilde{\sigma}_y \otimes \sigma_z,
\end{equation}
where $\tilde{\sigma}_x=g(\delta\theta)\sigma_x$, $\tilde{\sigma}_y=g(\delta\theta)\sigma_y$, and $g(\delta\theta)=\frac{2}{\delta\theta}\cos\!\left(\frac{\pi-\delta\theta}{2}\right)$. These can be regarded as approximate Pauli operations. The expectation value $\langle \tilde{W} \rangle=\mathrm{Tr}(\tilde{W}\rho)$ is negative if the two masses are entangled (see Appendix~\ref{app:witness}). Using Eq.\eqref{ent-state} with phases $\Delta \phi_{LR},\Delta \phi_{RL}$ from the discussion below Eq.\eqref{bose}, the expected witness measure at $\tau\sim 3\,$s is $\sim -0.0065$. Since each Pauli-correlator estimator has unit-order variance per shot, statistical resolution of this value at the $3\sigma$ level requires approximately $\mathcal{O}(10^5)$ experimental runs, which is comparable to the sample sizes envisaged for other proposed BMV implementations and is compatible with the $\sim 1\,$Hz decoherence-limited contrast discussed below.

\paragraph{Decoherence.}
During propagation, background gas collisions and black-body radiation unavoidably decohere the quantum coherence of the superposition state. Adopting the model in \cite{romero2011quantum}, air-molecule collisions decohere a $10^{-15}\,$kg test mass at a rate of $\sim 0.776\,$s$^{-1}$ under pressure $\sim 10^{-15}\,$Torr \cite{fitzakerley2016electron}. Black-body radiation arises from emission, absorption, and scattering of thermal photons. The emission localization parameter typically dominates because the internal temperature is usually larger than the external temperature. If the internal temperature can be cooled to $T_i=4\,$K, black-body radiation decoheres the test mass at a rate of $\sim 0.72\,$s$^{-1}$, corresponding to $\sim 10\,$s of coherence time. The required cooling to maintain the coherence of the superposition state is challenging with state-of-the-art technology.

\paragraph{Squeezing challenge.}
Taking an initial width $\sim 0.4\,$nm, the wavepacket expands to $\sim 0.85\,$nm after 3 seconds of free flight. We then have to squeeze the state by about 7 orders of magnitude to $1.5\times10^{-16}\,$m, so that it expands to $\sim 20\,\mu$m in the next $0.03\,$s, where interference occurs. Reaching this in the time budget requires diamagnetically levitated, sub-microhertz-dissipation oscillators with successive frequency jumps; the full implementation -- including the dissipation-rate constraint $\gamma<\hbar\omega_m^2/(n k_B T)$ for $n$ quarter-period jumps and an environmental temperature $T$ -- is given in Appendix~\ref{app:squeezing}.

\paragraph{Force noise.}
Beyond the squeezing-stage hardware noise, random forces present at any time $\delta F(t)$ give a decoherence rate
\begin{equation}
\Gamma \sim \frac{S_{FF}(\Omega) d^2}{\hbar^2},
\end{equation}
where $S_{FF}(\Omega)=\int \delta F(0)\delta F(t)e^{i\Omega t}dt$ is the force noise spectrum at the frequency $\Omega \sim 1/\tau$ of our experiment, and $d$ is the spatial splitting of each superposition. Keeping $\Gamma < 1\,$Hz gives the constraint $\sqrt{S_{FF}}\lesssim 10^{-29}\,$N$/\sqrt{\rm Hz}$. While this threshold appears demanding, faster-frequency noise does not particularly affect the experiment, while noise at this Hz frequency should be determinable by precision measurements over the long $\sim 1\,$s duration to be taken into account in the experiment. This requirement is not unique to the spatial qubit method; it is also present in Stern--Gerlach or Gaussian-state schemes. In fact, in those cases extra Stern--Gerlach inducing or trapping magnetic fields are perpetually present during the entire experimental run, whereas in the spatial qubit method we use free propagation for a large fraction of the time, when at least the randomness in the superposition-creating/delocalizing forces is inactive.

\section{Conclusions}
We have shown that a non-Gaussian, qubit-based witness of gravitationally induced entanglement between two $\sim 10^{-15}\,$kg masses can be implemented without ever embedding a spin in either mass. The two outstanding obstacles of the spin-based BMV scheme -- the demanding closure of a full Stern--Gerlach interferometer with simultaneous position and momentum overlap, and the additional decoherence channel introduced by the embedded spin -- are thereby removed in a single step. In their place, the protocol relies on the spreading of free quantum wavepackets to render the two localized branches $|L\rangle$ and $|R\rangle$ mutually overlapping at the interference plane, so that the qubit observables $\sigma_{x,y}=|L\rangle\langle R|\pm \text{h.c.}$ are directly read off from position-resolved detection events. A Faraday-shielded geometry that suppresses Casimir attraction between the two masses sets a minimum separation $d\sim 10\,\mu$m, which in turn fixes the new principal challenge of the scheme: the wavepacket must be squeezed by approximately seven orders of magnitude immediately after the gravitational entangling time $\tau$, so that the Pauli-$z$ and Pauli-$x,y$ measurements probe the \emph{same} entangled state of the two masses. The required squeezing is within the range that diamagnetically levitated micromechanical oscillators with sub-microhertz dissipation rates \cite{leng2021mechanical,zheng2020room,zhong2018towards} -- and, crucially, with in-situ frequency tunability of the trapping potential \cite{li2023diamagnetic} -- are now beginning to address, and we expect the consolidation of this single hardware capability to be the next experimental milestone on the path towards a BMV test. Subject to this and the standard coherence requirements (gas pressure, internal temperature), and assuming the required spatial superposition can be generated by either spinful or spinless routes (Appendix~\ref{app:initstate}), the conclusion stands: \emph{the quantum nature of gravity can be witnessed through position correlations alone, with no spin ever entering the readout}.

\begin{acknowledgments}
DH and US would like to acknowledge partial support from the DST-ITPAR grant IMT/Italy/ITPAR-IV/QP/2018/G. US also acknowledges partial support from the National Quantum Mission of DST. DH also acknowledges support from the NASI Senior Scientist Fellowship. AM's research is funded by the Netherlands Organization for Science and Research (NWO) grant number 680-91-119. BY acknowledges support from the National Natural Science Foundation of China (Grant No.~12404551) and the China Postdoctoral Science Foundation (Grant No.~2024M750339). SB acknowledges Grant No.~ST/W006227/1.
\end{acknowledgments}

\appendix

\section{Effective witness on separable states}\label{app:witness}

The construction of the witness in the main text requires the measured observables to follow the algebra of Pauli operations. The application of exact Pauli measurement would require arbitrarily small detection windows for $\sigma_x$ and $\sigma_y$, which is unphysical as no detection would take place. The applied operation using a finite-size detector is therefore an \emph{effective} witness:
\begin{equation}
\tilde{W}=I\otimes I-\tilde{\sigma}_x\otimes \tilde{\sigma}_x-\sigma_z\otimes \tilde{\sigma}_y-\tilde{\sigma}_y\otimes \sigma_z, \label{effW}
\end{equation}
where the error in the Pauli-$Z$ measurement is neglected due to its high efficacy (cf.\ \cite{yi2021massive}, Appendix A). The effective Pauli-$X$ and $Y$ measurements are $\tilde{\sigma}_x=g(\delta\theta)\sigma_x$ and $\tilde{\sigma}_y=g(\delta\theta)\sigma_y$, with $g(\delta\theta)=\frac{2}{\delta\theta}\cos\!\left(\frac{\pi-\delta\theta}{2}\right)$ (cf.\ \cite{yi2021massive}, Appendix C).

To show that the effective witness \eqref{effW} is valid, one needs to prove that it remains positive for all separable states. Consider a pure separable input state
\begin{eqnarray}
|\psi\rangle_1 \otimes|\psi\rangle_2 &=&\big(\cos(\alpha_1/2)|0\rangle+\sin(\alpha_1/2)e^{i\beta_1}|1\rangle\big)\nonumber\\
&&\otimes\big(\cos(\alpha_2/2)|0\rangle+\sin(\alpha_2/2)e^{i\beta_2}|1\rangle\big),\label{purestate}
\end{eqnarray}
where $\alpha,\beta\in[0,2\pi]$. If the witness is positive for all pure separable states, it is also positive for all separable states, as they are probability distributions over pure separable states.

Computing the expectation value of the effective witness with the pure input state \eqref{purestate} gives
\begin{eqnarray}
\langle \tilde{W} \rangle&=&1-\big[g(\delta\theta)^2\sin\alpha_1 \sin\alpha_2\cos\beta_1\cos\beta_2 \nonumber\\
&&+\, g(\delta\theta)\cos\alpha_1\sin\alpha_2\sin\beta_2 \nonumber\\
&&+\, g(\delta\theta)\cos\alpha_2\sin\alpha_1\sin\beta_1\big].
\end{eqnarray}
One can engineer the size of the detector to match the desired uncertainty in the phase angle $\delta\theta=\frac{md}{\hbar t_{\rm meas}^{x,y}}\delta x(y)$.

\begin{figure}[h!]
\centering
\includegraphics[width=0.85\columnwidth]{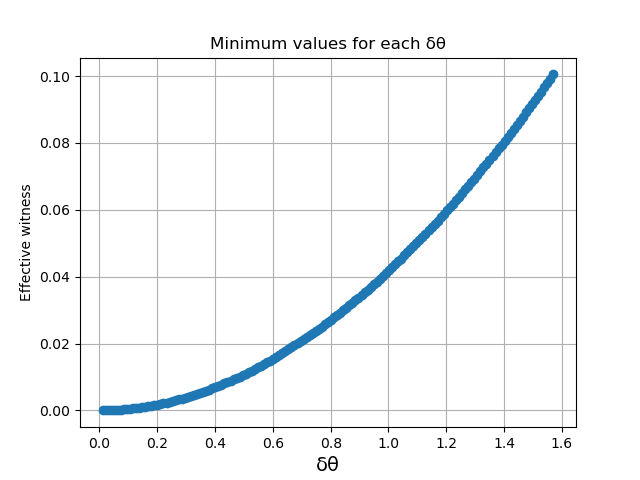}
\caption{\footnotesize \textbf{Minimum value of the effective witness over pure separable input states.} $\alpha_1,\alpha_2,\beta_1,\beta_2$, parametrising the pure input states, are each scanned in equal steps of 30 from 0 to $2\pi$. The minimum is plotted as a function of $\delta\theta$ (equal steps of 150).}
\label{effwit}
\end{figure}

Numerical simulation indicates that, generally, increasing $\delta\theta$ increases the expectation value of the effective witness. The minimum value of $\langle \tilde{W} \rangle$ over all pure separable input states is plotted in Fig.\ref{effwit} as a function of $\delta\theta$, where $\alpha$ and $\beta$ are scanned in $[0,2\pi]$ in equal steps of 30, and $\delta\theta\in[0,\pi/2]$ in equal steps of 150.

The plot indicates that the effective witness remains positive for all pure separable input states. If it returns a negative value, entanglement is then witnessed. Not all entangled states give rise to negative values; for the induced phase considered in the main text, the witness remains negative as long as the error in spatial detection is kept below $\delta\theta\sim 0.279$. Although the detector size smearing reduces the efficacy of the entanglement measurement, very good visibilities can still be obtained with proper care taken on the parameters; this spatial-qubit methodology has been implemented in quantum optics with finite-size detectors \cite{ghosh2022direct}.

\section{Initial state preparation}\label{app:initstate}

Witnessing the quantum nature of gravity through qubit correlations requires preparation of a superposition state of large spatial splitting $d\sim 10$--$100\,\mu$m of a micron-scale mass, $m\sim 10^{-15}$--$10^{-14}\,$kg \cite{bose2017spin,van2020quantum}. This is considerably beyond what has already been realized (e.g.\ a $10^{-25}\,$kg mass over $0.5\,$m \cite{kovachy2015quantum}, or a $10^{-22}\,$kg mass over $0.25\,\mu$m \cite{arndt2014testing,fein2019quantum}). However, several proposed schemes can achieve the required superpositions, with criteria in temperature, pressure, acceleration/vibration noise well identified.

The protocol starts from preparing a mass in a pure quantum state in a harmonic trap, typically the ground state. This mass may or may not have a spin embedded in it (the protocol of creating the initial state $\frac{1}{\sqrt{2}}(|L\rangle+|R\rangle)$ depends on that). We imagine these objects to be levitated in low-frequency static magnetic traps \cite{leng2021mechanical,zheng2020room,matsko2020mechanical,walker2019measurement}. Feedback cooling can be achieved by first shining light onto the particle. Scattered photons, which carry information about the particle's position, are collected by a photodiode detector. For example, from detecting $n$ scattered photons the position of a particle is determined to an accuracy of $\lambda/\sqrt{n}$, where $\lambda$ is the wavelength of the scattered light \cite{gieseler2012subkelvin}. This information is then used as feedback to cool the motion of the microsphere via an external damping force. Essentially, the information-gathering rate from detecting scattered photons must overtake the entropy increase rate of the object from undetected scatterings (photons, atoms) and noise from the environment. Using the feedback cooling principle, a mechanical oscillator as massive as 10\,kg has been prepared close to its ground state, the center-of-mass motion of which is cooled to tens of nanokelvin \cite{whittle2021approaching}, with position measurement uncertainty of $\sim 10^{-20}\,$m$/\sqrt{\rm Hz}$. Cryogenic diamagnetic levitated micromechanical oscillators have also been realized with very low ($\mu$Hz) dissipation rates \cite{leng2021mechanical,zheng2020room}. For a test mass $m\sim 10^{-15}\,$kg in a trap with frequency $\sim\mathcal{O}(10$--$100)\,$Hz, its ground-state spread is $\sim\mathcal{O}(0.1$--$1)\,$nm. As this uncertainty is much larger than the precision to which position has been localized in feedback cooling \cite{whittle2021approaching}, it is reasonable to suppose that feedback cooling to nearly the ground state is also imminent for the systems considered here (this will require, e.g., $n\sim 10^8$ scattered photons to be detected, with undetected photons being much fewer, in a time scale over which no collisions with air molecules take place). Following this stage, either spinful or spinless methods can be used to create the superposition.

If we have a spin embedded in the mass, a popular avenue to create a small superposition ($\leq 100\,$nm) is to use the Stern--Gerlach effect \cite{scala2013matter,yin2013large,wan2016free,bose2017spin,wood2022spin}. The test mass in an initially pure localized state $|C\rangle$ (say the ground state) is prepared with its spin in a state $\frac{1}{\sqrt{2}}(|\uparrow\rangle+|\downarrow\rangle)$ (by microwave pulses), and the trap is suddenly switched off. The spin state undergoes spatial splitting due to an inhomogeneous magnetic field gradient, so that the system evolves as
\begin{equation}
\tfrac{1}{\sqrt{2}}(|\uparrow\rangle+|\downarrow\rangle)|C\rangle \rightarrow \tfrac{1}{\sqrt{2}}(|L\uparrow\rangle+|R\downarrow\rangle).
\end{equation}
A subsequent measurement of the spin in a different basis, say the $|\pm\rangle$ basis, and getting the $|+\rangle$ outcome, prepares the mass in the spatial superposition
\begin{equation}
\tfrac{1}{\sqrt{2}}(|L\rangle+|R\rangle).
\end{equation}
Care must be taken so that the spin measurement does not reveal the position of the mass to a better precision than the $|L\rangle$--$|R\rangle$ separation.

We then need to amplify this superposition: in the case where there is electromagnetic screening, a spatial splitting of $\sim 10\,\mu$m is required, while the diamagnetism induced by the magnetic field gradient used for the splitting restricts the splitting \cite{pedernales2020motional,marshman2022constructing}. The superposition can be \emph{amplified} using a current-carrying wire providing a diamagnetism-induced repulsion between wire and each split component \cite{zhou2022mass}. Alternatively, spins may be subjected to nonlinear gradients to accumulate a velocity difference before catapulting to a large size \cite{zhou2022catapulting}. With magnetic field gradient $\sim 10^3\,$Tm$^{-1}$, the desired separation of $10\,\mu$m can be obtained after $\sim 1\,$s of flight time \cite{zhou2022catapulting}. During such intervals, the wavepacket spread remains in the $\sim\mathcal{O}(0.1$--$1)\,$nm regime. During these protocols, spin coherence does not need to be retained during the amplifications (diamagnetic repulsion/catapulting); the electronic spin does not play an active role. So, just before the amplification stage, one could map it to much more coherent nuclear spins, or simply measure it in a different basis as noted above to obtain directly the $\frac{1}{\sqrt{2}}(|L\rangle+|R\rangle)$ state. Moreover, for the application in the main text (using spatial qubits) we do not need to complete an interferometer -- only create the large-splitting superposition, which removes a significant challenge.

It is possible that even without spins the state $\frac{1}{\sqrt{2}}(|L\rangle+|R\rangle)$ can be produced as far as $\sim 1\,\mu$m-sized distances between $|L\rangle$ and $|R\rangle$ are concerned. For example, the first few stages of the on-chip interferometer of Ref.\cite{pino2018chip} based on coherent inflation and an $\hat{x}^2$ measurement can be used (until the superposition is generated). One can then, in principle, combine this with a diamagnetic-repulsion-aided further spatial splitting of the $|L\rangle$ and $|R\rangle$ terms \cite{zhou2022mass} so as to reach $\sim 10\,\mu$m size.

\section{Squeezing implementation: diamagnetic trapping and dissipation budget}\label{app:squeezing}

In the main text we noted that achieving the criterion $t^{x,y}_{\rm meas}\sim t^z_{\rm meas}=\tau$ requires position squeezing of the wavepacket by approximately seven orders of magnitude. Here we discuss the hardware implementation and dissipation budget.

Optical squeezing is the most common technique for microscopic objects. However, the momentum recoil due to photon scattering induces decoherence and heating on levitated objects, the rate of which scales with the object's size \cite{rahman2016burning,neumeier2022fast}. For a large mass $\sim 10^{-15}\,$kg this approach becomes infeasible: the resulting coherence time would be only $\sim 10^{-6}\,$s. Alternatively, one may adopt diamagnetic trapping for squeezing \cite{leng2021mechanical,zheng2020room,matsko2020mechanical,walker2019measurement}. Diamagnetic trapping typically operates at much lower frequencies than its optical counterpart, making it advantageous for holding large masses. The potential energy per unit volume of a diamagnetic mass trapped in a magnetic field reads
\begin{equation}
U \approx -\frac{\chi_m}{2\mu_0}B^2+\rho g r,
\end{equation}
where $\chi_m$ is the mass magnetic susceptibility, $\mu_0$ is the vacuum magnetic permeability, $g$ is the gravitational acceleration, $B$ is the induction of the magnetic field when the particle is absent, and $r$ is the vertical displacement.

The mechanical frequency is therefore
\begin{equation}
\omega_m=\sqrt{\frac{\chi_m}{\mu_0}}\,\frac{\partial B}{\partial r},
\end{equation}
where $\partial B/\partial r$ is the field gradient.

To control the motional superposition state during the squeezing procedure, the thermal decoherence rate $\gamma_{\rm th}=\bar{n}\gamma$ needs to be suppressed, where $\gamma$ is the mechanical dissipation rate and $\bar{n}=k_B T/(\hbar \omega_m)$ is the average phonon number. To keep coherence during $n$ quarter-period oscillations, the dissipation rate must satisfy
\begin{equation}
\gamma<\frac{\hbar \omega_m^2}{n k_B T}.
\end{equation}
To achieve the required squeezing, $n=7$ successive switches between two harmonic potentials of frequencies 100\,Hz and 1000\,Hz suffice. The environmental temperature can be kept at $\sim 10\,$mK with a commercial dilution refrigerator. The dissipation rate must then be kept below microhertz, $\gamma<1\,\mu$Hz.

For solid-state systems, the main contribution to dissipation is the direct coupling between the system and the substrate. Ultra-low dissipation in the microhertz range has been reported with diamagnetically levitated objects, where permanent magnets are used for trapping. In \cite{zheng2020room}, where the levitated mass is similar to our scheme, a damping rate of $\mu$Hz is achieved at pressure $\sim 3\times 10^{-7}\,$Torr and room temperature. The major contribution of damping comes from background gas collisions, which scales linearly with pressure. Dissipation due to gas collisions could be significantly reduced by lowering temperature and pressure. Comparing with the conditions to keep coherence during the propagation stage, one finds that the required level of vacuum is much less demanding during the squeezing stage.

However, a permanent-magnet-based scheme may be unfeasible for squeezing, since successive changes of trapping potential are required. Frequency-adjustable diamagnetic levitation, in which the trap frequency is tuned in situ during operation, has recently been demonstrated on the same Du-group platform used for the sub-microhertz dissipation measurements \cite{li2023diamagnetic}; the closely related capability of dynamically scanning parameter loops on a levitated micro-oscillator has also been realized \cite{yin2020chiral}, indicating that the successive frequency switches called for here are within experimental reach. Alternatively, one may use the magnetic field generated by current-carrying coils, which introduces an additional type of dissipation due to field fluctuations. Specific current sources must be considered, but we do not go into that technicality here. The overall constraint on broadband random force noise -- which is also relevant outside the squeezing stage -- is given in the main text (force-noise paragraph), and is not unique to the spatial-qubit method.

\bibliographystyle{apsrev4-2}
\bibliography{ref}

\end{document}